\documentclass{article}

\usepackage{arxiv}

\usepackage[utf8]{inputenc}
\usepackage[T1]{fontenc}
\usepackage{hyperref}
\usepackage{url}
\usepackage{booktabs}
\usepackage{amsfonts}
\usepackage{amsmath}
\usepackage{nicefrac}
\usepackage{microtype}
\usepackage{threeparttable}
\usepackage{textgreek}
\usepackage{lipsum}
\usepackage{graphicx}
\usepackage[justification=centering]{caption}
\usepackage{subcaption}

\usepackage{xcolor}
\usepackage{soul}

\usepackage{setspace} 
\usepackage{parskip}  

\graphicspath{ {./images/} }

\sethlcolor{yellow} 

\title{Reduced Molecular Plasma Kinetics with Non-Boltzmann Vibrational Distribution}

\usepackage{fancyhdr}

\author{
Sagar Pokharel$^{*}$ \\
Dept. of Aerospace Engineering and Eng. Mechanics\\
The University of Texas at Austin\\
Austin, TX 78712\\
\texttt{spokharel@utexas.edu}\\
$^{*}$Corresponding author\\
\And
Laxminarayan L. Raja\\
Dept. of Aerospace Engineering and Eng. Mechanics\\
The University of Texas at Austin\\
Austin, TX 78712\\
\texttt{lraja@mail.utexas.edu}\\
}

\begin{document}

\maketitle
\begin{abstract}

Vibrational kinetics can strongly affect energy transfer, chemical kinetics, and discharge behavior in molecular plasmas, but resolving all internal states is often impractical for reactor-scale simulation. Compact, physics-based vibrational closures are particularly needed when complete state-to-state (StS) data are limited. Starting from a detailed \(\mathrm{H}_2\) StS plasma-kinetics framework coupled to an electron-Boltzmann calculation, this work develops a reduced vibrational-kinetics formulation. The \(\mathrm{H}_2\) vibrational ladder is reduced to two non-Boltzmann groups that evolve their population and internal energy. State-resolved reaction rates and associated energy-exchange terms are projected onto the groups for subsequent tabulation in reactor-scale plasma models. StS-informed samples of the reduced-model are propagated for pulsed RF \(\mathrm{H}_2\) plasmas across a scan of pulse-averaged absorbed power and compared with the corresponding StS simulations. Predictions from the sampled models reproduce the bulk plasma response, including electron density, atomic-H density, and \(\mathrm{H}_3^+\) density. Discrepancies in \(\mathrm{H}^-\) production and the upper-energy-group vibrational temperature persist across the sampled reductions. This indicates a limitation of the present VDF reconstruction rather than the nominal parameter choice. Tail-sensitive processes therefore require a more accurate representation of the vibrational distribution.

\end{abstract}

\section{Introduction}
\label{sec:introduction}

Low-temperature molecular plasmas are studied for a range of energy and aerospace-relevant applications. They are widely used for material processing in semiconductor manufacturing \cite{economouPulsedPlasmaEtching2014}, and they provide a route to plasma-assisted chemical conversion through selective electron-driven molecular activation \cite{bogaertsPlasmaTechnologyCO22020}. Electric-propulsion systems have conventionally relied on noble-gas propellants \cite{choueiriPhysicsPlasmaPropulsion2011,goebelFundamentalsElectricPropulsion2008}; however, alternative concepts based on molecular feedstocks, including air-breathing propulsion for very-low-Earth-orbit operation, motivate the study of molecular plasma kinetics in a propulsion setting \cite{tagawaExperimentalStudyAir2013}. Pulsed Radio-frequency (RF) operation introduces temporal control beyond continuous-wave excitation and has been used to modify plasma properties in both processing and high-density discharges \cite{economouPulsedPlasmaEtching2014,bannaPulsedHighdensityPlasmas2012}.The pulse frequency and duty cycle set the relative durations of the ignition, sustained-discharge, and afterglow phases, and thereby the balance among electron-impact processes \cite{bruneauEffectGasProperties2015,hernandezElectronDynamicsReignition2020,pokharelNonEquilibriumElectronStatetoState2026}.

Vibrational excitation is a major internal-energy channel in molecular plasmas. Vibrationally excited molecules can modify electron-impact dissociation, attachment, and ionization pathways, while vibrational--translational (V--T) transfer and wall interactions thermalize the stored energy \cite{capitelliVibrationalKineticsElectron2006,colonnaRoleMolecularVibration2016}. The vibrational-state distribution also remains coupled to the electron energy distribution function (EEDF) through inelastic excitation and superelastic collisions, particularly under low-field and post-discharge conditions \cite{colonnaRoleMolecularVibration2016,capitelliSelfconsistentApproachCold2024}. The relevant kinetics depend on how the vibrational energy is distributed across the molecular levels. The effect is strongest when a long-lived vibrational population feeds a channel that is weak from the ground state. In H$_2$, for example, dissociative attachment increases by orders of magnitude with vibrational level, so H$^-$ formation is governed by the vibrationally excited population \cite{pokharelStatetostateKineticsCoupled2026}. Vibrational energy is transferred to the gas through V--T relaxation and can be deposited at surfaces through wall de-excitation, affecting gas temperature and surface heat loads \cite{tsonevSimulationPulsedCO22025,piskinEUVinducedLowPressure2025}. 

State-to-state (StS) models resolve these effects by retaining the population of each internal level, whether vibrational, rovibrational, or across the coupled modes of polyatomic molecules such as CO$_2$. The number of levels and the need for level-resolved kinetic data limit their direct use in multidimensional reactor simulations. For many molecular mixtures and operating conditions, the required level-resolved data are incomplete or unavailable. A common reduction lumps levels into groups and evolves moments of their populations. Panesi and coworkers formulated this as a general multigroup method, in which each group's population and energy are evolved and the reconstructed intra-group distribution defines the grouped rate and energy-transfer terms \cite{liuGeneralMultigroupMacroscopic2015,sharmaCoarsegrainedModelingThermochemical2020}. For N$_2$+N, two or three vibrational groups recovered bulk dissociation where a single vibrational temperature underpredicted it \cite{munafoModelingDissociationEnergy2015}. In CO$_2$ microwave plasmas, Berthelot and Bogaerts showed that few-group models can reproduce bulk plasma variables while misrepresenting the vibrational tail that controls dissociation \cite{berthelotModelingPlasmabasedCO22016}.

At reactor scale, non-equilibrium plasmas are commonly represented by multi-temperature models that solve separate energy equations for the electrons, the gas, and a bulk vibrational mode. This approach has been validated against measured electron and gas temperatures in N$_2$--O$_2$ plasmas \cite{pokharelSelfconsistentModelNumerical2023b,urdanetaTemporallyResolvedProperties2025}. Recent work has moved toward embedding reduced closures derived from StS kinetics in such self-consistent and multidimensional models \cite{kumarInvestigationNonequilibriumPhenomena2024,colonnaMultitemperatureModelStatespecific2026,jeongFirstPrincipleInspiredReducedOrderModels2026}. These StS-derived closures are valuable when detailed kinetic information is available, but their construction remains contingent on that information. For a pulsed CO$_2$ glow discharge, Tsonev \emph{et al.} replaced a state-resolved scheme with a few vibrational-energy balances coupled to an electron Boltzmann solver, reproducing measured vibrational- and gas-temperature transients \cite{tsonevSimulationPulsedCO22025}. Such closures typically assume a Boltzmann distribution within each group or mode. In reactor-scale frameworks, electron-impact rate coefficients are also commonly supplied as functions of a single electron temperature, with a prescribed energy-exchange term for each reaction. A useful reduction must therefore retain the vibrational-distribution dependence that materially changes these rates and energy exchanges, while remaining compatible with practical reactor-scale simulations and providing a path for cases in which the StS information is limited.

Our earlier work developed a coupled unsteady electron--Boltzmann and StS model for RF H$_2$ plasmas that resolves fifteen vibrational levels together with the EEDF and plasma composition \cite{pokharelStatetostateKineticsCoupled2026}. The normalized distribution of molecular population across these vibrational levels is referred to here as the vibrational distribution function (VDF). Over a range of pulsed conditions, the VDF was strongly non-Boltzmann but could be represented by piecewise-Boltzmann segments \cite{pokharelNonEquilibriumElectronStatetoState2026}. These resolved calculations provide the reference solutions and VDF data for the present reduction. Building on this , the present work develops and assesses a two-group moment-based reduction for pulsed H$_2$ plasmas. The group populations and internal energies are evolved, while a constrained reconstruction retains non-Boltzmann VDF structure within each group. State-resolved electron-impact and heavy-particle processes are projected onto the groups to define reduced rate and energy-exchange closures. The resulting representation provides a compact internal-state model that can be tabulated or further parameterized for reactor-scale plasma simulations.

The reduced model is compared with the StS calculations over a pulsed power scan at \(100~\mathrm{mTorr}\). The reference VDFs also define StS-informed samples of the reduced-model parameterization. Propagating these samples assesses the prediction spread introduced by the reduction and its closure representation, in addition to the agreement of the nominal reduced model with the StS reference.

The remainder of the paper first describes the vibrational state resolved H$_2$ cases and VDF segmentation. It then presents the moment-based group formulation, reconstruction closure, projected kinetic terms, and rate representation. The reduced and StS calculations are then compared, the sensitivity to the nominal segment temperatures is analyzed, and the conclusions are presented.


\section{Reference ensemble and VDF structure}
\label{sec:reference_vdf}

\subsection{Reference VDF ensemble and segment selection}
\label{sec:reference_vdf_ensemble}

The VDF representation used below is established from the coupled electron-Boltzmann and StS H$_2$ plasma-kinetics model developed in our earlier works~\cite{pokharelStatetostateKineticsCoupled2026, pokharelNonEquilibriumElectronStatetoState2026}. The spatially homogeneous model resolves the populations of H$_2(v=0\text{--}14)$, the EEDF, and the coupled heavy-species kinetics. We use pulsed-steady VDFs from 28 operating conditions at \(100~\mathrm{mTorr}\), spanning absorbed power, duty cycle, and pulse frequency; this set is referred to as the \emph{reference ensemble}. These data are used to identify a reduced VDF structure that can provide vibrational-state information to offline electron-Boltzmann calculations and support the reduced plasma-kinetics formulation without evolving the full StS system.

\begin{figure}[t]
\centering
\includegraphics[width=0.45\linewidth]{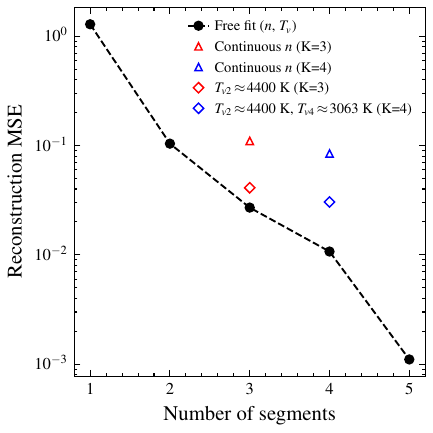}
\caption{VDF reconstruction error over the reference ensemble versus the number of segments, \(K\). Filled circles show the minimum-error unconstrained fits with independent segment populations and temperatures; open triangles enforce continuity across all segment boundaries. Diamonds show fixed-temperature reconstructions with pairwise continuity. For the selected \(K=4\) representation, with cuts after \(v=1\), \(5\), and \(12\), continuity is imposed between the \(v=0\text{--}1\) and \(v=2\text{--}5\) segments, and separately between the \(v=6\text{--}12\) and \(v=13\text{--}14\) segments; the latter segment in each pair has a fixed temperature.}
\label{fig:vdf_segmentation}
\end{figure}

The resolved VDFs are strongly non-Boltzmann and exhibit a depleted tail associated with dissociation processes \cite{pokharelNonEquilibriumElectronStatetoState2026}. We therefore represent the distribution using \(K\) piecewise-Boltzmann segments, each with a local vibrational temperature, \(T_{v,k}\), that defines the relative populations over its assigned range of levels. The number of segments was varied from \(K=1\) to \(5\), with the cuts for each \(K\) selected to minimize the mean-squared error in logarithmic population over the reference ensemble. As shown by the solid circles in Fig.~\ref{fig:vdf_segmentation}, the minimum reconstruction error decreases as segments are added. The selected four-segment representation, \(\mathcal{S}_1:v=0\text{--}1\), \(\mathcal{S}_2:v=2\text{--}5\), \(\mathcal{S}_3:v=6\text{--}12\), and \(\mathcal{S}_4:v=13\text{--}14\), isolates the dissociation-depleted tail. Although a fifth segment further reduces the fitting error, it adds another independently parameterized portion of the distribution and its associated collisional-kinetics treatment.

The prior multi-temperature analysis in Ref.~\cite{pokharelNonEquilibriumElectronStatetoState2026} also showed that selected portions of the VDF retain comparatively stable slopes over the studied operating conditions. We therefore retain \(T_{v,2}\) and \(T_{v,4}\) as nominal fixed temperatures (more details in the following sections) and use the four segments to define two vibrational groups, \(G_1:v=0\text{--}5\) and \(G_2:v=6\text{--}14\). The reduced plasma-kinetics model evolves the population and internal energy of each group. Within each group, continuity is imposed between \(\mathcal{S}_1\)--\(\mathcal{S}_2\) and \(\mathcal{S}_3\)--\(\mathcal{S}_4\), respectively, so that the population of each fixed-temperature segment follows from its adjacent dynamic segment. The diamond markers in Fig.~\ref{fig:vdf_segmentation} show the reconstruction error associated with this pairwise representation. 

Under the local-energy approximation, electron-impact rates in reactor-scale models are commonly obtained from an offline electron-Boltzmann calculation when reliable collision cross sections are available; otherwise, literature-based reaction-rate expressions are used. Including vibrational non-equilibrium requires the VDF as an additional input. A fully unconstrained piecewise representation would require an independent population and temperature for every segment, producing a high-dimensional rate database that is practical only when adequate state-resolved data are available and unwieldy for a general reduced framework. We instead impose continuity across all segment boundaries, so that the total H$_2(v)$ population and segment temperatures determine the VDF while solving for the EEDF. The reference VDFs do not exhibit strong population inversion, supporting this continuous, monotone parameterization; its reconstruction errors are shown by the open triangles in Fig.~\ref{fig:vdf_segmentation}.

\subsection{Transient VDF evolution}
\label{sec:fixed_temperature_closure}

\begin{figure}
\centering
\begin{subfigure}[b]{0.43\textwidth}
\centering
\includegraphics[width=\textwidth]{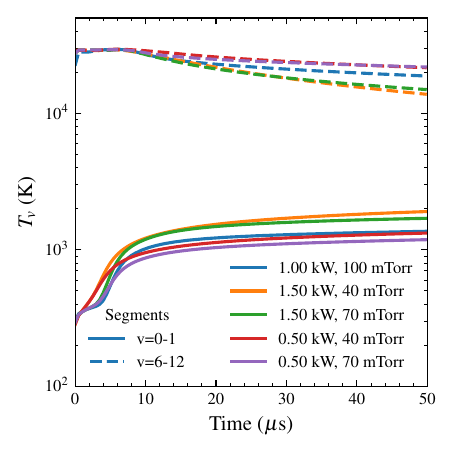}
\caption{\(T_{v,1}\) and \(T_{v,3}\).}
\label{fig:transient_01}
\end{subfigure}
\hfill
\begin{subfigure}[b]{0.43\textwidth}
\centering
\includegraphics[width=\textwidth]{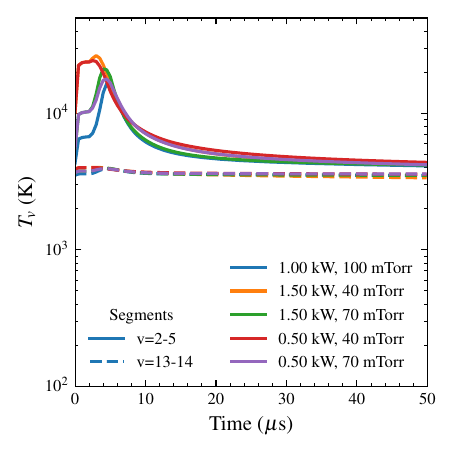}
\caption{\(T_{v,2}\) and \(T_{v,4}\).}
\label{fig:transient_02}
\end{subfigure}
\caption{Cold-start evolution of the fitted segment temperatures using the four-segment cuts selected in Fig.~\ref{fig:vdf_segmentation}. The \(T_{v,2}\) and \(T_{v,4}\) segments exhibit an early overshoot before relaxing to their post-ignition values, whereas \(T_{v,1}\) and \(T_{v,3}\) retain slower evolution.}
\label{fig:segment_transient}
\end{figure}

The segmentation and reconstruction errors in Fig.~\ref{fig:vdf_segmentation} are determined from steady VDFs. We therefore apply the same \(K=4\) representation to the initial evolution of several cold-start CW cases, initialized at low electron density and spanning a range of power and pressure. Figure~\ref{fig:segment_transient} shows that \(T_{v,2}\) and \(T_{v,4}\) undergo a pronounced early overshoot before relaxing and converging within approximately \(15~\mu\mathrm{s}\). In contrast, \(T_{v,1}\) and \(T_{v,3}\) do not exhibit comparable convergence and evolve gradually over the transient. This supports using \(T_{v,2}\) and \(T_{v,4}\) as fixed nominal temperatures after the initial ignition transient. This assumption is appropriate for sustained discharges whose residence times are much longer than \(15~\mu\mathrm{s}\).

\subsection{Two-group VDF reconstruction}
\label{sec:two_group_formulation}

\begin{figure}
\centering
\includegraphics[width=0.60\linewidth]{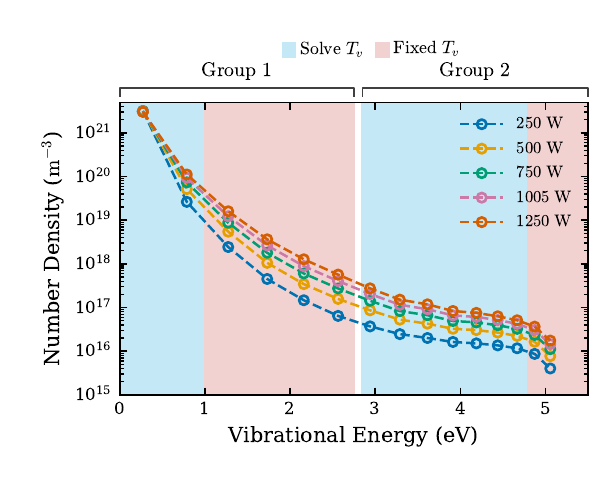}
\caption{Two-group VDF representation used in the reduced model. Group \(G_1\) comprises the dynamic low-energy segment \(\mathcal{S}_1\) and the fixed-temperature segment \(\mathcal{S}_2\); group \(G_2\) similarly comprises the dynamic segment \(\mathcal{S}_3\) and fixed-temperature tail segment \(\mathcal{S}_4\).}
\label{fig:two_group_structure}
\end{figure}

The selected \(K=4\) representation is reduced to two vibrational groups, \(G_1=\mathcal{S}_1\cup\mathcal{S}_2\), spanning \(v=0\text{--}5\), and \(G_2=\mathcal{S}_3\cup\mathcal{S}_4\), spanning \(v=6\text{--}14\), as illustrated in Fig.~\ref{fig:two_group_structure}. The reduced plasma-kinetics model evolves the population and internal energy of each group. Within \(G_1\), \(T_{v,1}\) remains dynamic while \(T_{v,2}\) is fixed; within \(G_2\), \(T_{v,3}\) remains dynamic while \(T_{v,4}\) is fixed. Continuity within each segment pair then determines the population of the fixed-temperature segment from its adjacent dynamic segment. No continuity is imposed between \(G_1\) and \(G_2\), allowing their populations and internal energies to evolve independently from there governing equations.



The following formulation evolves these group moments and reconstructs the level populations required to evaluate the projected kinetic source terms.

\section{Moment-based group formulation for the reduced-model kinetics}
\label{sec:moment_based}

\subsection{Group kinetics formalism}

The detailed StS model provides the high-dimensional kinetic baseline from which grouped rate and energy-transfer closures are projected. The moment-based reconstruction used here follows established multi-group formulations for thermochemical nonequilibrium kinetics~\cite{liuGeneralMultigroupMacroscopic2015,munafoModelingDissociationEnergy2015}.

For a molecular species \(A\), let \(A_i\) denote internal state \(i\), with population \(n_i\) and internal energy \(\varepsilon_i\). Its state-resolved evolution is governed by Eq.~\eqref{eq:sts_master}, where \(\Omega_i\) is the net production rate of state \(i\), \(\mathbf{n}\) is the full StS population vector, and \(\eta\) denotes the quantities required to evaluate the microscopic kinetics, such as the reduced electric field (or EEDF) or gas temperature.

\begin{equation}
    \label{eq:sts_master}
    \frac{d n_i}{dt} = \Omega_i(\mathbf{n};\eta),
\end{equation}

\begin{equation}
    \label{eq:group_moment_projection}
    \frac{dN_g}{dt} = S^N_g(\mathbf{n}^{\star};\eta),
    \qquad
    \frac{dU_g}{dt} = S^U_g(\mathbf{n}^{\star};\eta),
\end{equation}

\begin{equation}
    \label{eq:group_projected_sources}
    S^N_g(\mathbf{n}^{\star};\eta) = \sum_{i\in\mathbb{A}_g} \Omega_i(\mathbf{n}^{\star};\eta),
    \qquad
    S^U_g(\mathbf{n}^{\star};\eta) = \sum_{i\in\mathbb{A}_g} \varepsilon_i \Omega_i(\mathbf{n}^{\star};\eta).
\end{equation}

\begin{equation}
    \label{eq:group_reconstruction}
    n^{\star}_i = N_g p_{i|g}(\chi_g), 
    \qquad 
    \sum_{i\in\mathbb{A}_g} p_{i|g}(\chi_g) = 1, 
    \qquad 
    e_g = \frac{U_g}{N_g} = \sum_{i\in\mathbb{A}_g} \varepsilon_i p_{i|g}(\chi_g).
\end{equation}

The internal states are partitioned into disjoint macro-groups \(\mathbb{A}_g\). For each group, the reduced variables are the group population, \(N_g=\sum_{i\in\mathbb{A}_g}n_i\), and group internal energy, \(U_g=\sum_{i\in\mathbb{A}_g}\varepsilon_i n_i=N_g e_g\), where \(e_g\) is the group mean internal energy, $\epsilon$ is the state-level energy. Taking the zeroth and first energy moments of Eq.~\eqref{eq:sts_master} yields the grouped evolution equations in Eq.~\eqref{eq:group_moment_projection}. The population and energy source terms are obtained by summing the state-resolved source terms over the corresponding group, as defined in Eq.~\eqref{eq:group_projected_sources} and the source terms evaluated using a reconstructed population vector \(\mathbf{n}^{\star}\). Within each group, the reconstruction is defined by Eq.~\eqref{eq:group_reconstruction}, where \(p_{i|g}(\chi_g)\) is a normalized intra-group distribution and \(\chi_g\) denotes its closure parameter or parameters.

With the exact StS reconstruction, \(\mathbf{n}^{\star}=\mathbf{n}\), Eqs.~\eqref{eq:group_moment_projection} and \eqref{eq:group_projected_sources} give the exact zeroth and first energy moment projection of Eq.~\eqref{eq:sts_master}. The reduced-model approximation therefore enters through the reconstruction of \(\mathbf{n}^{\star}\), and hence through the intra-group distribution \(p_{i|g}\). The formulation requires a normalized distribution \(p_{i|g}(\chi_g)\) for which the mean-energy relation is invertible, so that \(\chi_g\) is uniquely determined by the evolved group mean energy \(e_g\). A conventional, but not unique, choice is the local Boltzmann distribution in Eq.~\eqref{eq:local_boltzmann}, parametrized by the group temperature \(T_g\) (for vibrational excitation considered here, \(g_i=1\)).

\begin{equation}
\label{eq:local_boltzmann}
    p_{i|g}(T_g)
    =
    \frac{g_i\exp(-\varepsilon_i/(k_B T_g))}
    {Z_g(T_g)},
    \qquad
    Z_g(T_g)
    =
    \sum_{j\in\mathbb{A}_g}
    g_j\exp(-\varepsilon_j/(k_B T_g)).
\end{equation}

\subsubsection{Single-Parameter Non-Boltzmann Closure}
\label{subsec:single_parameter_non_boltzmann_closure}

Within each two-segment group defined in Section~\ref{sec:two_group_formulation}, let \(\mathbb{A}_g^{(1)}\) and \(\mathbb{A}_g^{(2)}\) denote the dynamic (lower-energy) and fixed-temperature (higher-energy) segments, respectively. The dynamic inverse temperature is \(\beta=(k_B T_g)^{-1}\), while the fixed inverse temperature is \(\beta_{\mathrm{fix}}=(k_B T_{\mathrm{fix}})^{-1}\). The within-group reconstruction is defined by Eq.~\eqref{eq:piecewise_distribution_beta}. The coefficients in Eq.~\eqref{eq:piecewise_distribution_beta} are defined by Eq.~\eqref{eq:piecewise_ai_Ci_definition}, where \(E_c\) is the joining energy between the two segments.

\begin{equation}
    \label{eq:piecewise_distribution_beta}
    p_{i|g}(\beta)
    =
    \frac{C_i\exp(-\beta a_i)}
    {Z_g(\beta)},
    \qquad
    Z_g(\beta)
    =
    \sum_{j\in\mathbb{A}_g}
    C_j\exp(-\beta a_j).
\end{equation}

\begin{equation}
    \label{eq:piecewise_ai_Ci_definition}
    a_i
    =
    \begin{cases}
        \varepsilon_i, & i\in\mathbb{A}_g^{(1)},\\
        E_c, & i\in\mathbb{A}_g^{(2)},
    \end{cases}
    \qquad
    C_i
    =
    \begin{cases}
        g_i, & i\in\mathbb{A}_g^{(1)},\\
        g_i\exp[-\beta_{\mathrm{fix}}(\varepsilon_i-E_c)],
        & i\in\mathbb{A}_g^{(2)}.
    \end{cases}
\end{equation}

Equation~\eqref{eq:piecewise_distribution_beta} retains the dynamic slope in \(\mathbb{A}_g^{(1)}\) and the prescribed slope in \(\mathbb{A}_g^{(2)}\), with continuity enforced at \(E_c\). Once \(T_{\mathrm{fix}}\) is prescribed, \(\beta\), or equivalently \(T_g\), is the single closure parameter and defines the intra-group distribution \(p_{i|g}\). The required invertibility of the map from \(T_g\) to the group mean energy, \(e_g\), and hence the uniqueness of the reconstruction, is established in Appendix~\ref{app:piecewise_closure_invertibility}.

\begin{figure}[t]
    \centering
    \includegraphics[width=0.4\textwidth]{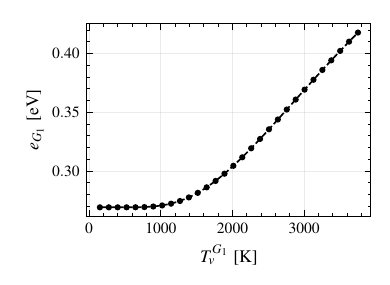}
    \hspace{0.02\textwidth}
    \includegraphics[width=0.4\textwidth]{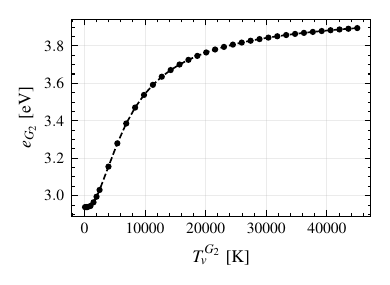}
    \caption{Group mean internal vibrational energy as a function of the dynamic group vibrational temperature for the two-group reconstruction. Left: \(e_{G_1}(T_v^{G_1})\), with \(T_{v,2}^{0}\approx4400~\mathrm{K}\). Right: \(e_{G_2}(T_v^{G_2})\), with \(T_{v,4}^{0}\approx3100~\mathrm{K}\).}
    \label{fig:group_energy_temperature_maps}
\end{figure}

For the two-group representation in Section~\ref{sec:two_group_formulation}, the dynamic group temperatures \(T_v^{G_1}\) and \(T_v^{G_2}\) determine the corresponding group mean energies through the within-group reconstruction in Eq.~\eqref{eq:group_reconstruction}. Figure~\ref{fig:group_energy_temperature_maps} shows these relations with \(T_{v,2}^{0}\approx4400~\mathrm{K}\) and \(T_{v,4}^{0}\approx3100~\mathrm{K}\) held fixed. Their monotonicity allows the evolved group mean energies to be uniquely inverted to obtain the dynamic closure temperatures.

\subsubsection{Population-Weighted Grouped Rates and Energy Exchange}
\label{subsec:effective_grouped_rates}

The grouped moment equations require each state-resolved reaction family to be represented by a reaction rate and by the associated signed internal-energy transfers. These quantities are obtained by population-weighting the microscopic channels with the reconstructed reactant distributions. The construction is written for a general process involving two internally resolved molecules of species \(A\) and an ungrouped collision partner \(b\); reactions with fewer internally resolved reactants or products follow by omitting the unused state and group indices.

The state-resolved process and its corresponding grouped reaction are shown in Eq.~\eqref{eq:grouped_reaction_pair}.

\begin{equation}
    \label{eq:grouped_reaction_pair}
    A_i+A_j+b
    \rightarrow
    A_m+A_n+b,
    \quad
    \substack{
        i\in\mathbb{A}_g,\;
        j\in\mathbb{A}_h,\\
        m\in\mathbb{A}_p,\;
        n\in\mathbb{A}_q}
    \qquad\Longrightarrow\qquad
    \mathbb{A}_g+\mathbb{A}_h+b
    \rightarrow
    \mathbb{A}_p+\mathbb{A}_q+b.
\end{equation}

Here, \(b\) may be an electron, neutral bath species, or ion, and \(\rho\equiv gh\rightarrow pq\) denotes the grouped reaction channel. Given the state-resolved rate coefficient \(k_{ijmn}(\eta)\), the grouped rate coefficient and macroreaction rate are defined by Eq.~\eqref{eq:grouped_rate_coefficient}.

\begin{equation}
    \label{eq:grouped_rate_coefficient}
    \bar{k}_{\rho}(\boldsymbol{\chi};\eta)
    =
    \sum_{i\in\mathbb{A}_g}
    \sum_{j\in\mathbb{A}_h}
    p_{i|g}p_{j|h}
    \sum_{m\in\mathbb{A}_p}
    \sum_{n\in\mathbb{A}_q}
    k_{ijmn}(\eta),
    \qquad
    R_{\rho}
    =
    n_b N_gN_h\bar{k}_{\rho}.
\end{equation}

\begin{equation}
    \label{eq:rate_weighted_average}
    \left\langle X\right\rangle_{\rho}
    =
    \frac{1}{\bar{k}_{\rho}}
    \sum_{i\in\mathbb{A}_g}
    \sum_{j\in\mathbb{A}_h}
    p_{i|g}p_{j|h}
    \sum_{m\in\mathbb{A}_p}
    \sum_{n\in\mathbb{A}_q}
    k_{ijmn}X_{ijmn}.
\end{equation}

The energy terms are rate-normalized averages over the same microscopic channels. For any state-resolved event quantity \(X_{ijmn}\), this average is defined by Eq.~\eqref{eq:rate_weighted_average}. For each participating group \(r\in\{g,h,p,q\}\), Eq.~\eqref{eq:net_evib} defines the signed vibrational-energy contribution of reaction \(\rho\) per macroreaction event. A positive value deposits vibrational energy into group \(r\), whereas a negative value removes it. The corresponding population and internal-energy source contributions are given by Eq.~\eqref{eq:grouped_moment_contributions}.

\begin{equation}
    \label{eq:net_evib}
    \Delta\varepsilon^{\mathrm{vib}}_{r,\rho}
    =
    \left\langle
        \delta_{rp}\varepsilon_m
        +
        \delta_{rq}\varepsilon_n
        -
        \delta_{rg}\varepsilon_i
        -
        \delta_{rh}\varepsilon_j
    \right\rangle_{\rho}.
\end{equation}

\begin{equation}
    \label{eq:grouped_moment_contributions}
    S_r^N\mathrel{+}=\nu_{r,\rho}R_\rho,
    \qquad
    S_r^U\mathrel{+}
    =
    R_\rho\Delta\varepsilon^{\mathrm{vib}}_{r,\rho},
    \qquad
    \nu_{r,\rho}
    =
    \delta_{rp}+\delta_{rq}-\delta_{rg}-\delta_{rh}.
\end{equation}

\begin{equation}
    \label{eq:grouped_collision_loss}
    Q^{\mathrm{eff}}_{b,\rho}
    =
    \left\langle \ell_b\right\rangle_{\rho},
    \qquad
    P_{b,\rho}
    =
    R_{\rho}Q^{\mathrm{eff}}_{b,\rho}.
\end{equation}

The energy exchange with the ungrouped collision partner \(b\) is projected using the same rate weighting. Let \(\ell_{b,ijmn}(\eta)\) denote the signed energy lost by \(b\) per occurrence of the microscopic channel \(i,j\rightarrow m,n\), with \(\ell_{b,ijmn}>0\) when energy is removed from \(b\). Equation~\eqref{eq:grouped_collision_loss} defines the corresponding rate-normalized energy loss per macroreaction event, \(Q^{\mathrm{eff}}_{b,\rho}\), and its associated power loss, \(P_{b,\rho}\).

Thus, each grouped reaction channel carries the effective rate coefficient \(\bar{k}_{\rho}\), the signed vibrational-energy contributions \(\Delta\varepsilon^{\mathrm{vib}}_{r,\rho}\) for its participating groups, and the collision-partner energy loss \(Q^{\mathrm{eff}}_{b,\rho}\). These projected quantities depend directly on the reconstructed reactant vibrational population distributions \(p_{i|g}(\chi_g)\) and \(p_{j|h}(\chi_h)\), and on the microscopic kinetic parameters \(\eta\). When electron-impact quantities are obtained from a VDF-conditioned electron-Boltzmann calculation, \(\eta\) also inherits some dependence on the complete reconstructed VDF (check later section on this).

\subsubsection{Parameter Bounds for High and Low Collision Energies}
\label{subsec:parameter_selection_high_low_te}

When detailed state-resolved kinetic data are unavailable, limiting estimates can still be constructed for the signed group-energy contribution \(\Delta\varepsilon^{\mathrm{vib}}_{r,\rho}\) in Eq.~\eqref{eq:net_evib}. Consider a channel \(\rho\) that consumes a molecule from group \(\mathbb{A}_g\) and produces no vibrationally resolved molecular product, as in direct dissociation. In this case, \(\Delta\varepsilon^{\mathrm{vib}}_{g,\rho}=-\langle\varepsilon_i\rangle_\rho\). To estimate this quantity, we assume two limiting forms for the state-dependent reaction rate. In the low-collision-energy limit, we assume that only the highest-energy state in the group contributes to the reaction. In the high-collision-energy limit, we assume that all states react with the same rate coefficient. Equation~\eqref{eq:group_energy_limits} gives the resulting estimates.

\begin{equation}
    \label{eq:group_energy_limits}
    \Delta\varepsilon^{\mathrm{vib}}_{g,\rho}
    \simeq
    \begin{cases}
        -\varepsilon_{\mathrm{top},g},
        & \text{low collision energy},\\
        -e_g,
        & \text{high collision energy},
    \end{cases}
    \qquad
    \varepsilon_{\mathrm{top},g}
    =
    \max_{i\in\mathbb{A}_g}\varepsilon_i.
\end{equation}

\begin{equation}
    \label{eq:qeff_constant_residual}
    Q^{\mathrm{eff}}_{b,\rho}
    \simeq
    E^{0}_{b,\rho}
    +
    \sum_r
    \Delta\varepsilon^{\mathrm{vib}}_{r,\rho}.
\end{equation}
For the state-resolved reaction in Eq.~\eqref{eq:grouped_reaction_pair}, \(Q^{\mathrm{eff}}_{b,\rho}\) can be estimated when the energy loss of collision partner \(b\) satisfies \(\ell_{b,ijmn}\simeq E^{0}_{b,\rho}+\varepsilon_m+\varepsilon_n-\varepsilon_i-\varepsilon_j\), where \(E^{0}_{b,\rho}\) is independent of the microscopic states. Population weighting this relation gives Eq.~\eqref{eq:qeff_constant_residual}. Direct electron-impact dissociation through the \(\mathrm{b}^3\Sigma_\mathrm{u}^+\) repulsive state satisfies \(\ell_{e,i}=D_0-(\varepsilon_i-\varepsilon_0)\), where \(D_0\) is the threshold from \(v=0\) and \(\varepsilon_0\) is its zero-point energy. Thus, \(E^{0}_{e,\rho}=D_0+\varepsilon_0\) \cite{pokharelNonEquilibriumElectronStatetoState2026}. For this channel, the threshold decreases monotonically along the vibrational ladder. For a resonant vibrational transition \(i \rightarrow m\), \(\ell_{e,im}=\varepsilon_m-\varepsilon_i\), so \(E^{0}_{e,\rho}=0\). Equations~\eqref{eq:group_energy_limits} and \eqref{eq:qeff_constant_residual} therefore provide limiting estimates for the group-energy contribution and \(Q^{\mathrm{eff}}_{e,\rho}\), respectively, when state-resolved data are unavailable. These assumptions are reaction channel dependent. In the present work, the available StS data allow their applicability and the resulting limiting estimates to be assessed directly.

\begin{figure}[t]
   \centering
   \includegraphics[width=\linewidth]{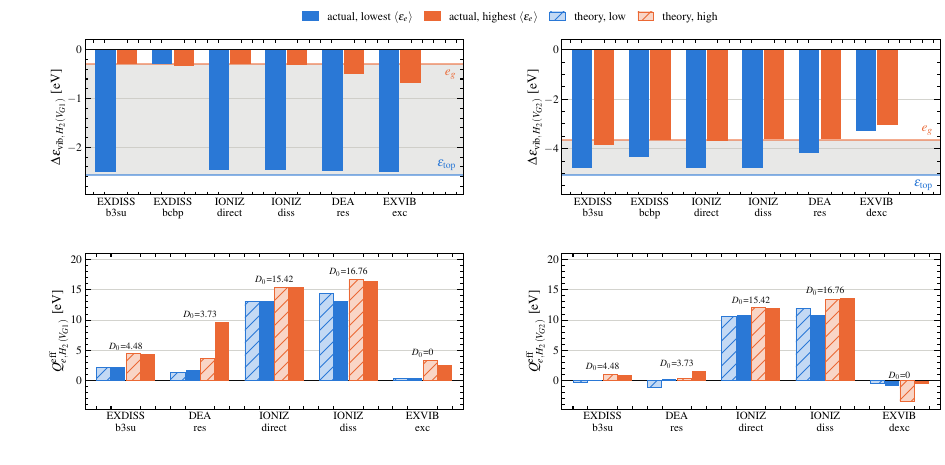}
   \caption{Comparison of StS-projected vibrational-energy and electron-energy quantities with the limiting estimates of Eqs.~\eqref{eq:group_energy_limits} and \eqref{eq:qeff_constant_residual}. The top panels show \(\Delta\varepsilon^{\mathrm{vib}}_{g,\rho}\), and the bottom panels show \(Q^{\mathrm{eff}}_{e,\rho}\). The gray range denotes the limiting interval for \(\Delta\varepsilon^{\mathrm{vib}}_{g,\rho}\).}
   \label{fig:bounds_4panel}
\end{figure}

Figure~\ref{fig:bounds_4panel} shows that the estimates capture the vibrational-energy contribution for dissociation through the \(\mathrm{b}^3\Sigma_\mathrm{u}^+\) repulsive state (DISS b3su), direct ionization (IONIZ direct), and dissociative ionization (IONIZ diss), and give a reasonable estimate for vibrational excitation (EXVIB exc). The associated \(Q^{\mathrm{eff}}_{e,\rho}\) estimate is accurate when the vibrational-energy contribution is captured and a state-independent \(E^{0}_{e,\rho}\) exists. The main discrepancies reflect violations of these conditions. Dissociation through the singlet \(B^1\Sigma_\mathrm{u}^+\), \(C^1\Pi_\mathrm{u}\), and \(B'^1\Sigma_\mathrm{u}^+\) states (DISS bcbp) has a nonmonotonic threshold, so the highest vibrational level is not the dominant low-energy reactant \cite{pokharelStatetostateKineticsCoupled2026}. Vibrational de-excitation (EXVIB dexc) is instead dominated by levels near the group boundary. The \(Q^{\mathrm{eff}}_{e,\rho}\) estimate is less reliable for the singlet-state dissociation and dissociative-ionization channels because no single \(E^{0}_{e,\rho}\) represents their microscopic processes. It is also less reliable for resonant dissociative attachment (DEA res), for which the exported electron-energy term is not threshold based.

\subsection{EEDF Solution and Reduced-Kinetics Database}
\label{subsec:eedf_reduced_kinetics}
The collision-kinetics rate database is constructed offline using the coupled electron-Boltzmann and plasma-kinetics framework developed previously~\cite{pokharelStatetostateKineticsCoupled2026,pokharelNonEquilibriumElectronStatetoState2026}, which resolves the EEDF using detailed collision cross sections. The calculations are performed for pure \(\mathrm{H}_2\) at \(100~\mathrm{mTorr}\), with \(10~\mathrm{MHz}\) RF excitation and \(T_{\mathrm{gas}}=300~\mathrm{K}\). The EEDF calculation uses the globally continuous four-segment VDF defined above, with \(T_{v,2}^{0}\approx4400~\mathrm{K}\) and \(T_{v,4}^{0}\approx3100~\mathrm{K}\) held fixed, while \(T_{v,1}\) and \(T_{v,3}\) are varied over \(300\text{--}2500~\mathrm{K}\) and \(5000\text{--}30000~\mathrm{K}\), respectively. These VDF conditions are combined with imposed electric fields between \(0.1\) and \(5000~\mathrm{V\,m^{-1}}\) to define the offline scan. For each condition, the VDF-defined \(\mathrm{H}_2\) population distribution is held fixed while the electron-Boltzmann solver is advanced to a periodic-steady state and cycle-averaged to obtain the EEDF, state-resolved electron-impact rate coefficients, and associated electron-energy-loss terms.

\begin{figure}[t]
    \centering
    \includegraphics[width=0.6\textwidth]{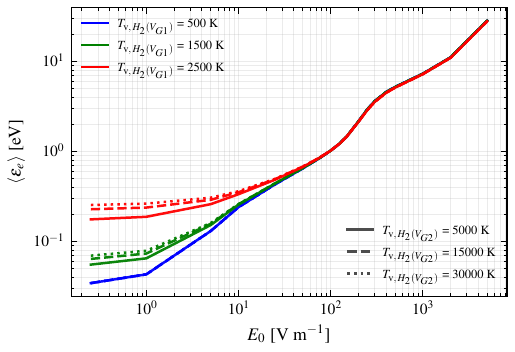}
    \caption{Mean electron energy, \(\langle\varepsilon_e\rangle\), as a function of imposed electric field for representative VDF conditions. The imposed field does not uniquely determine \(\langle\varepsilon_e\rangle\) when the VDF changes at low electric field.}
    \label{fig:mean_electron_energy_vdf_scan}
\end{figure}

\begin{figure}[t]
    \centering
    \includegraphics[width=0.85\textwidth]{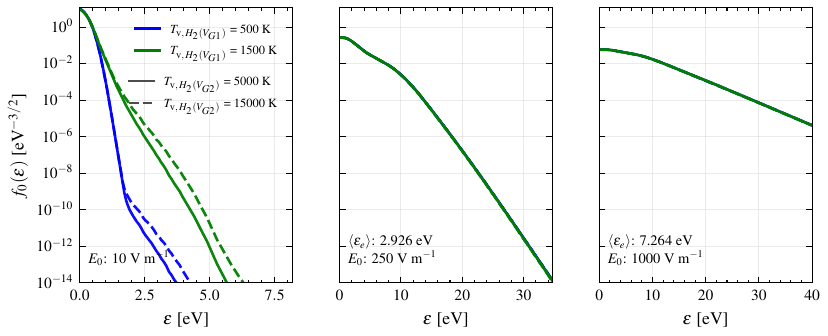}
    \caption{Representative cycle-averaged EEDFs for selected VDF conditions, illustrating the dependence of the electron-energy distribution on the vibrational population.}
    \label{fig:eedf_vdf_comparison}
\end{figure}

At a fixed electric field, changes in the prescribed VDF modify the EEDF and therefore the mean electron energy \(\langle\varepsilon_e\rangle\). This effect is most pronounced at low electric field, where superelastic collisions with vibrationally excited \(\mathrm{H}_2\) provide an appreciable electron-heating pathway. At the fixed operating condition above, the electron-impact quantities are therefore parameterized by \(\langle\varepsilon_e\rangle\) and the four VDF segment temperatures \((T_{v,1},T_{v,2},T_{v,3},T_{v,4})\). Figures~\ref{fig:mean_electron_energy_vdf_scan} and \ref{fig:eedf_vdf_comparison} illustrate the sensitivity of the mean electron energy and EEDF to the VDF parameters.

The state-resolved kinetic quantities and collision-partner energy-loss terms obtained from the offline scan are population-weighted using the reconstructed group distributions and projected according to Eqs.~\eqref{eq:grouped_rate_coefficient}, \eqref{eq:net_evib}, and \eqref{eq:grouped_collision_loss}. Each grouped reaction channel is consequently characterized by an effective rate coefficient \(\bar{k}_{\rho}\), signed group-energy contributions \(\Delta\varepsilon^{\mathrm{vib}}_{r,\rho}\) for the participating vibrational groups, and an effective energy loss \(Q^{\mathrm{eff}}_{b,\rho}\) for collision partner \(b\). 

\subsection{Univariate Reduction of the Kinetic Database}
\label{subsec:univariate_database_reduction}

Retaining the full multivariate dependence of every projected quantity on \((\langle\varepsilon_e\rangle,T_v^{G_1},T_v^{G_2})\), where \(T_v^{G_1}\equiv T_{v,1}\) and \(T_v^{G_2}\equiv T_{v,3}\), is not practical for implementation in general reactor-scale solvers. Moreover, without an StS reference calculation, this dependence cannot generally be established. The reduction therefore seeks a minimal representation of the vibrational kinetics that remains compatible with existing reactor-scale modeling frameworks. The two-group projection reduces approximately \(10^3\) state-resolved collision processes to 74 grouped channels covering electron-impact and heavy-particle interactions.

For each electron-impact grouped channel, a one-dimensional rate profile is fitted independently against \(\langle\varepsilon_e\rangle\), \(T_v^{G_1}\), and \(T_v^{G_2}\). Each candidate is scored by the fraction of the total variation in \(\ln\bar{k}_{\rho}\) reproduced by its profile, and the coordinate with the largest score is retained as the channel's tabulation coordinate. In the present database, \(\langle\varepsilon_e\rangle\) is selected for 48 of the 74 channels, \(T_v^{G_1}\) for 10, and \(T_v^{G_2}\) for 8; the remaining 8 channels have negligible variation and are exported as scalar rate constants.

For electron-impact channels, the group-energy contributions \(\Delta\varepsilon^{\mathrm{vib}}_{r,\rho}\) and electron-energy losses \(Q^{\mathrm{eff}}_{e,\rho}\) are exported as rate-weighted scalar constants. For either quantity \(q_\rho\), the exported value is \(q_\rho^{\mathrm{scalar}}=\langle\bar{k}_{\rho}q_\rho\rangle/\langle\bar{k}_{\rho}\rangle\), where the averages are taken from \(\varepsilon_{\mathrm{low},\rho}\) to \(5~\mathrm{eV}\), with \(\bar{k}_{\rho}(\varepsilon_{\mathrm{low},\rho})=0.01\,\bar{k}_{\rho}(5~\mathrm{eV})\). For reactions with threshold energy and retaining the \(\langle\varepsilon_e\rangle\)-axis as tabulation coordinate, the exported rate is set to zero below its low-energy cutoff. The reaction and its associated energy source are then inactive at sufficiently low electron energy.

Representative profiles for resonant dissociative attachment, direct dissociation, ionization, and resonant electron-vibrational transfer are provided in Appendix~\ref{app:group_resolved_channel_profiles}. The gray traces show the projected quantities over representative VDF and electron energy conditions, while the magenta rate profile and dashed scalar energy terms give the reduced representation at the nominal VDF condition. Their spread illustrates the VDF dependence not retained by the single-coordinate rate and scalar energy closures, motivating the model reduction related sensitivity analysis below.

\section{Assessment of the Reduced Model in Pulsed RF Plasmas}
\label{sec:pulsed_steady_assessment}

The reduced model is compared with the detailed StS model for the same pulsed-RF \(\mathrm{H}_2\) discharges over an applied-power scan. Apart from the treatment of the vibrational ladder, both models use the same reactor conditions and power-control procedure.

The simulations represent a cylindrical inductively coupled plasma reactor with a diameter of \(37~\mathrm{cm}\) and a height of \(15~\mathrm{cm}\). Pure \(\mathrm{H}_2\) is supplied at \(100~\mathrm{mTorr}\), with a gas temperature of \(300~\mathrm{K}\) and a flow residence time of \(1~\mathrm{ms}\). Power is supplied by a \(10~\mathrm{MHz}\) RF source. Here, the discharge is pulsed at \(10~\mathrm{kHz}\) with a duty cycle of 0.5, and the pulse-averaged absorbed power is varied from \(250\) to \(1250~\mathrm{W}\). For each case, a controller adjusts the electric-field amplitude to maintain the target absorbed power. Both models are advanced until a pulse-periodic steady state is reached, and all reported quantities are cycle averaged. Details of the implementation and the StS response over this power range were reported previously \cite{pokharelNonEquilibriumElectronStatetoState2026}. The comparisons below assess whether the reduced model reproduces the sustained-discharge behavior and associated vibrational kinetics.

\subsection{StS-Informed Sampling of VDF Parameterizations}
\label{subsec:closure_sensitivity}

\begin{figure}[t]
    \centering
    \includegraphics[width=0.865\textwidth]{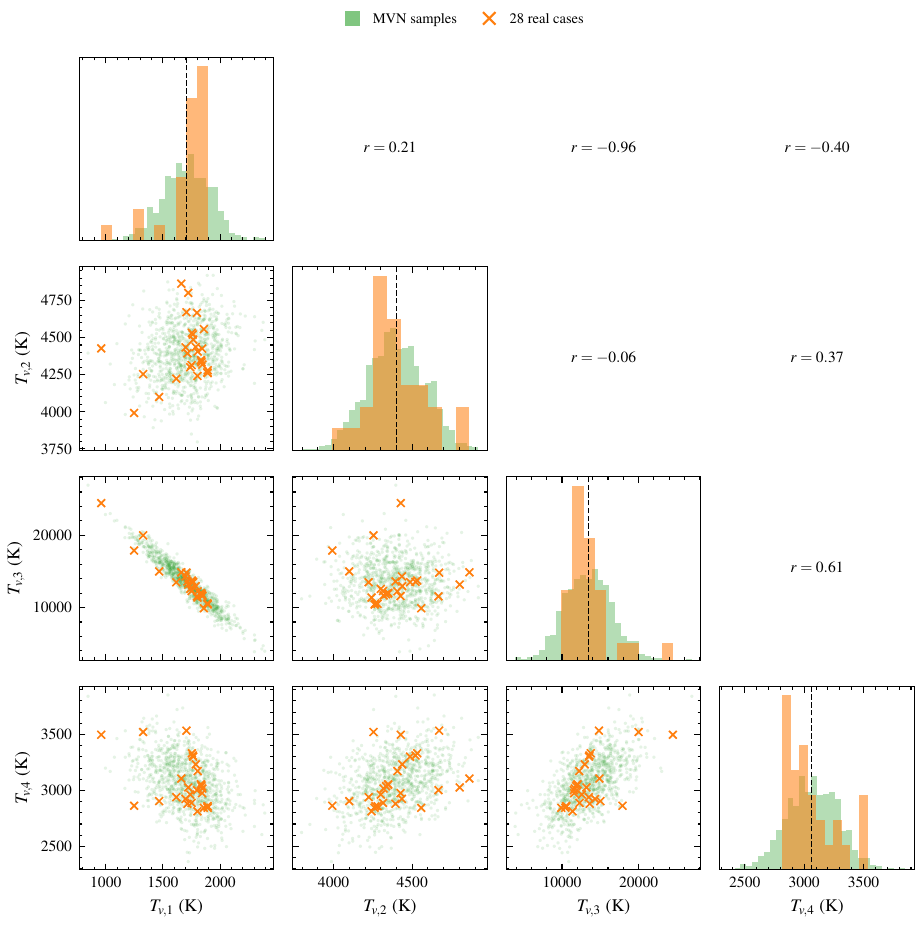}
    \caption{Joint distribution of the four segment temperatures fitted to the reference ensemble. The mean defines the nominal reduced-model parameterization. The off-diagonal panels show the correlations retained when sampling alternative parameter sets.}
    \label{fig:closure_point_distribution}
\end{figure}

Previous uncertainty analyses of plasma models have focused on uncertain kinetic inputs or unresolved reactor-scale physics. Berthelot and Bogaerts propagated uncertain rate coefficients through a detailed CO$_2$ plasma-chemistry model and found that vibrational populations and conversion can be substantially more uncertain than bulk electron properties \cite{berthelotModelingCO2Plasma2017}. Barberena-Valencia \textit{et al.}\ treated unresolved reactor-scale physics as uncertain parameters in a reduced CCP model and propagated their effect on the predictions \cite{barberena-valenciaProbabilisticZerodimensionalModel2025}. The present analysis is complementary. It does not vary the StS kinetic data or calibrate the reduced model to the StS solution. Instead, it examines the structural sensitivity of the reduced VDF representation and the resulting uncertainty in predictions introduced by the reduced model construction.

The nominal reduced database uses a segment-temperature vector, \(\mathbf{T}_v=(T_{v,1},T_{v,2},T_{v,3},T_{v,4})^\mathsf{T}\), to set the fixed upper-segment temperatures and the VDF-coordinate values omitted from the univariate rate and scalar energy-exchange representations. This is an engineering compromise that avoids retaining the full multi-coordinate dependence of kinetic parameters. To assess its sensitivity, \(\mathbf{T}_v\) is sampled from a multivariate-normal distribution fitted to the 28-case reference ensemble. Its mean, \(\boldsymbol{\mu}\), defines the nominal database, while its covariance matrix, \(\boldsymbol{\Sigma}\), retains the observed correlations among the fitted segment temperatures. Figure~\ref{fig:closure_point_distribution} shows the reference fits and 1000 correlated draws from the fitted multivariate-normal distribution. In particular, \(T_{v,1}\) and \(T_{v,4}\) exhibit a strong negative correlation.

For each draw, a separate reduced reaction-mechanism is constructed and propagated through the same pulsed-steady calculations. The sampled vector changes the projected rate and energy-exchange terms through the fixed reconstruction parameters and omitted-coordinate values. The EEDF database retains its dependence on the dynamic temperatures, \(T_{v,1}\) and \(T_{v,3}\), but is not recomputed for sampled values of the fixed upper-energy-segment temperatures, \(T_{v,2}\) and \(T_{v,4}\). The fixed temperatures have a comparatively small fitted spread, while the more highly populated lower-energy segments dominate the low-field VDF--EEDF coupling (see Fig. \ref{fig:eedf_vdf_comparison}). This retains the leading EEDF response without constructing a separate EEDF database for every draw. The scalar electron-impact energy terms retain the rate-weighted mean-electron-energy reduction of Section~\ref{subsec:univariate_database_reduction}. The resulting spread measures the sensitivity of the reduced closure database under this EEDF approximation.

\subsection{Comparison Against the StS Results}

The following figures compare the StS and reduced-model results. The shaded central \(50\%\) and \(95\%\) intervals show the spread obtained from the sampled parameterizations described earlier.

\begin{figure}[t]
    \centering
    \includegraphics[width=0.48\textwidth]{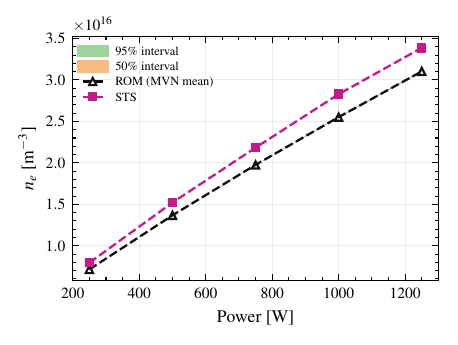}
    \hfill
    \includegraphics[width=0.48\textwidth]{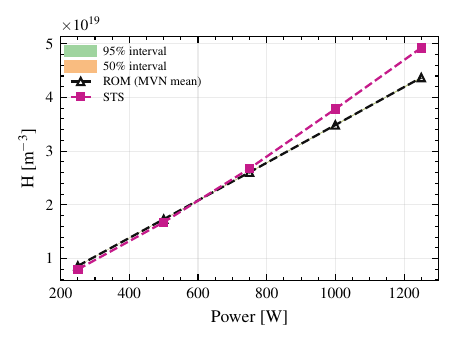}
    \caption{Electron density \(n_e\) (left) and total atomic-H density (right) as functions of applied power.}
    \label{fig:sensitivity_ne_Htotal}
\end{figure}

\begin{figure}[t]
    \centering
    \includegraphics[width=0.48\textwidth]{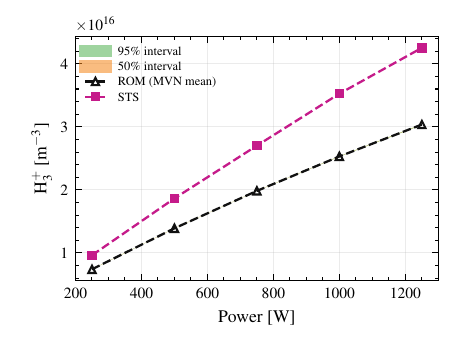}
    \hfill
    \includegraphics[width=0.48\textwidth]{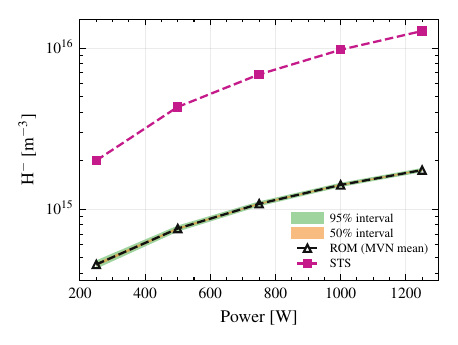}
    \caption{\(\mathrm{H}_3^+\) density (left) and \(\mathrm{H}^-\) density (right) as functions of applied power.}
    \label{fig:sensitivity_h3plus_hminus}
\end{figure}

Figure~\ref{fig:sensitivity_ne_Htotal} shows that the electron density and total atomic-H density are only weakly affected by the sampled parameterizations. In both panels, the nominal reduced-model result remains close to the StS result across the power scan, and the sampled intervals are negligible on the plotted scale. Figure~\ref{fig:sensitivity_h3plus_hminus} shows the corresponding charged-species comparison. The \(\mathrm{H}_3^+\) density remains similarly robust, whereas \(\mathrm{H}^-\) is underpredicted by a factor of \(4\text{--}7\). Its sampled range is only \(6\text{--}12\%\) of the median value, so this discrepancy is not caused by the sampled reduced-model parameterization.

\begin{figure}[t]
    \centering
    \includegraphics[width=0.48\textwidth]{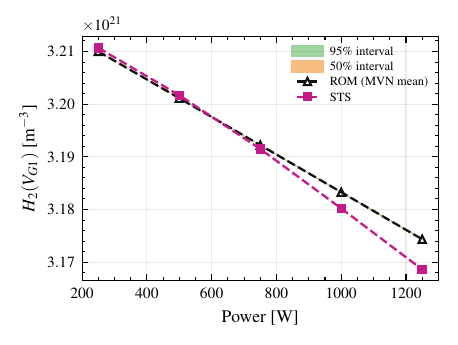}
    \hfill
    \includegraphics[width=0.48\textwidth]{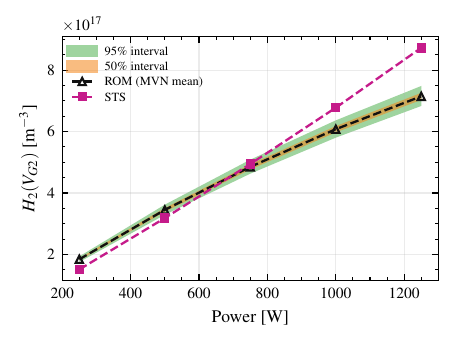}
    \caption{Populations of \(\mathrm{H}_2(V_{G_1})\) (left) and \(\mathrm{H}_2(V_{G_2})\) (right) as functions of applied power.}
    \label{fig:sensitivity_n_G1_G2}
\end{figure}

\begin{figure}[t]
    \centering
    \includegraphics[width=0.48\textwidth]{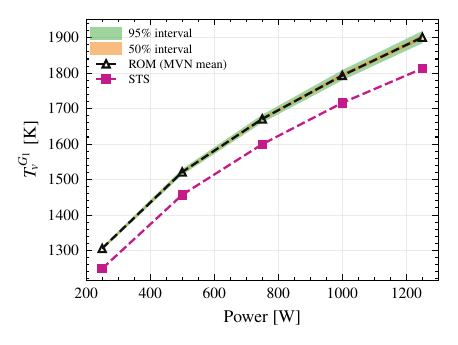}
    \hfill
    \includegraphics[width=0.48\textwidth]{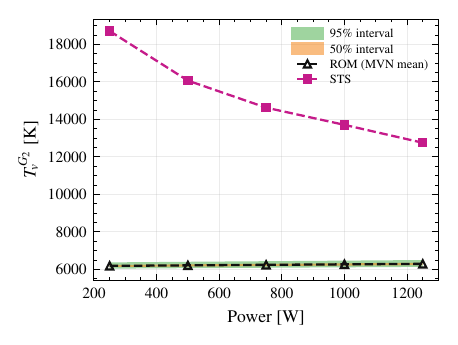}
    \caption{Group vibrational temperatures \(T_v^{G_1}\) (left) and \(T_v^{G_2}\) (right) as functions of applied power.}
    \label{fig:sensitivity_Tvib_G1_G2}
\end{figure}

\begin{figure}[t]
    \centering
    \includegraphics[width=0.48\textwidth]{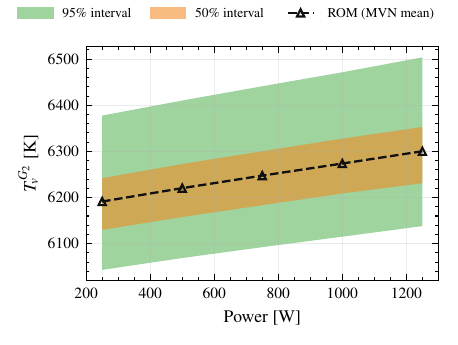}
    \caption{Reduced-model \(T_v^{G_2}\) over the sampled parameterizations. The StS result is omitted to resolve the sampled spread; the full comparison is shown in Fig.~\ref{fig:sensitivity_Tvib_G1_G2}.}
    \label{fig:sensitivity_Tvib_G2_zoom}
\end{figure}

The group-resolved quantities in Fig.~\ref{fig:sensitivity_n_G1_G2} show different behavior in the two groups. The \(\mathrm{H}_2(V_{G_1})\) population remains close to the StS result and is essentially unchanged by the sampled parameterizations. For \(\mathrm{H}_2(V_{G_2})\), the sampled range is comparable to the StS discrepancy at low power but becomes too small to account for the difference as power increases. Figure~\ref{fig:sensitivity_Tvib_G1_G2} shows the corresponding group vibrational temperatures. The \(T_v^{G_1}\) discrepancy remains small, whereas \(T_v^{G_2}\) differs from the StS result by a factor of \(2\text{--}3\). The zoomed result in Fig.~\ref{fig:sensitivity_Tvib_G2_zoom} shows that the sampled variation in \(T_v^{G_2}\) is only approximately \(5\text{--}6\%\). The reduced model also does not capture the decreasing trend of \(T_v^{G_2}\) with power, and the sampled spread is too small to account for this difference.

The remaining discrepancies identify limits of the present reduction. The representative rate profiles in Appendix~\ref{app:group_resolved_channel_profiles} show that the rates, group-energy contributions \(\Delta\varepsilon^{\mathrm{vib}}_{r,\rho}\), and electron-energy losses \(Q^{\mathrm{eff}}_{e,\rho}\) retain dependence on both \(\langle\varepsilon_e\rangle\) and the VDF parameters. In contrast, the reduced model retains univariate rate representations and rate-weighted scalar energy terms. The sampled parameterizations modify these reduced quantities, but the resulting spread is too small to explain the \(\mathrm{H}^-\) and \(T_v^{G_2}\) discrepancies. These discrepancies should be interpreted together because the dissociative electron attachment (DEA) channels couple \(\mathrm{H}^-\) formation directly to the vibrational population.

The \(T_v^{G_2}\) discrepancy is better interpreted as a population-distribution, or shape, limitation than as a group-population error. The total \(\mathrm{H}_2(V_{G_2})\) population agrees with the StS result to within \(22\%\), and the corresponding mean-energy difference is only \(6\text{--}9\%\), although the fitted group temperatures differ by a factor of \(2\text{--}3\). The dominant excitation channels populate levels near the lower edge of \(G_2\), whereas subsequent vibration--vibration transfer in the StS model can redistribute this energy toward higher vibrational levels and sustain a non-Boltzmann tail. The present two-segment reconstruction cannot vary this tail independently of the lower part of the group. These effects provide a plausible structural explanation for the \(T_v^{G_2}\) discrepancy and, through the vibrational dependence of the DEA channels, may also contribute to the \(\mathrm{H}^-\) shortfall. Nevertheless, the model remains stable over the sampled parameter range and provides robust predictions for the bulk plasma response.

\subsection{Implications and Limits of the Present Reduction}

The comparison establishes both a useful operating range and a clear limitation of the present reduction. The two-group model reproduces the bulk plasma response across the power scan, including the electron density, atomic-H density, \(\mathrm{H}_3^+\) density, and lower-group population. In contrast, the persistent \(\mathrm{H}^-\) and \(T_v^{G_2}\) discrepancies are not explained by the sampled parameterizations. They indicate that similar group population and mean energy do not necessarily preserve the high-energy distribution required by tail-sensitive processes.

This distinction between robust bulk predictions and errors in tail-dependent kinetics is consistent with earlier reduced plasma models \cite{berthelotModelingPlasmabasedCO22016}. The present results therefore identify both the utility and the limit of the compact closure. For \(\mathrm{H}_2\), the high-vibrational tail affects dissociative attachment and the resulting negative-ion production. The small spread under the sampled parameterizations shows that this behavior cannot be corrected by adjusting the assumed segment temperatures within the present reconstruction. Instead, it requires a richer representation that can follow the VDF more closely. Because the pulsed-steady state develops over repeated pulses, a persistent error in the VDF-tail can accumulate over successive cycles.

\section{Conclusions}
\label{sec:conclusions}

This work developed and assessed a reduced vibrational-kinetics formulation for pulsed \(\mathrm{H}_2\) plasmas. The vibrational ladder was represented by two groups that evolve their population and internal energy. A four-segment piecewise-Boltzmann reconstruction retained the observed non-Boltzmann VDF structure without evolving the full state-to-state system. Within each group, the lower-energy segment evolves with the group energy, while the upper-energy segment is fixed from the StS reference ensemble.

State-resolved reaction rates and event-conditioned vibrational and electron-energy-transfer quantities were population-weighted and projected into two vibrating groups. The rate database was further reduced to univariate representations, and the corresponding energy-transfer terms were represented by rate-weighted scalar constants. This provides the internal-state reduction step needed before the resulting closures are tabulated or further parameterized for a particular reactor-scale plasma model.

Comparison with the pulsed-steady StS calculations showed that the reduced model remained stable and reproduced the bulk plasma response across the applied-power scan. Electron density, atomic-H density, \(\mathrm{H}_3^+\) density, and the lower-group population were robust to the sampled reduced-model parameterizations. In contrast, the \(\mathrm{H}^-\) density and upper-group vibrational temperature were not recovered accurately. Their narrow sampled ranges show that these discrepancies arise from the present VDF representation rather than from the nominal fixed-temperature and coordinate-collapse choices. The comparison identifies a clear limit of the two-group closure. Tail-sensitive processes, including vibrationally enhanced dissociative attachment, require a richer representation of the upper vibrational distribution.

More broadly, the present formulation complements established StS-informed multigroup reductions. It provides a compact non-Boltzmann VDF reconstruction and distribution-conditioned rate and energy-transfer closures for reactor-scale implementations. When detailed kinetic information is incomplete, the proposed limiting estimates provide physically motivated approximations for unresolved energy-transfer terms. This framework is relevant to molecular plasmas with mature internal-state models and to complex gases such as \(\mathrm{CF}_4\), \(\mathrm{NF}_3\), and \(\mathrm{SF}_6\), for which complete vibrational state-to-state data remain limited. The present \(\mathrm{H}_2\) results provide a detailed benchmark for this strategy.

\section*{Acknowledgement}
This work was supported by the Air Force Office of Scientific Research under Grant No.\ FA9550-24-1-0358 (Program Manager: Dr.\ Justin Koo).

\appendix
\section*{Appendix}
\section{Uniqueness of the Two-Segment Group VDF Reconstruction}
\label{app:piecewise_closure_invertibility}

This result applies to the two-segment closures introduced in Section~\ref{subsec:single_parameter_non_boltzmann_closure}, in which a fixed-temperature upper segment is joined continuously to a dynamic lower segment. For fixed \(\beta_{\mathrm{fix}}\), Eq.~\eqref{eq:piecewise_distribution_beta} defines a one-parameter exponential-family distribution, with the fixed upper-segment slope absorbed into \(C_i\). Let \(\langle q\rangle_g=\sum_{i\in\mathbb{A}_g}q_i p_{i|g}\) denote the within-group average. Differentiating the energy-consistency relation in Eq.~\eqref{eq:group_reconstruction} with respect to \(\beta\) gives Eq.~\eqref{eq:energy_derivative_covariance}.

\begin{equation}
    \label{eq:energy_derivative_covariance}
    \frac{\partial e_g}{\partial\beta}
    =
    \sum_{i\in\mathbb{A}_g}
    \varepsilon_i p_{i|g}
    \left(\langle a\rangle_g-a_i\right)
    =
    e_g\langle a\rangle_g
    -
    \langle\varepsilon a\rangle_g
    =
    -\operatorname{Cov}_g(\varepsilon,a).
\end{equation}

The quantity \(a_i\) defined by Eq.~\eqref{eq:piecewise_ai_Ci_definition} is non-decreasing with \(\varepsilon_i\). Chebyshev's sum inequality therefore gives \(\operatorname{Cov}_g(\varepsilon,a)\geq0\)~\cite{hardyInequalities2001}. For the closures used here, each dynamic lower segment contains states with \(\varepsilon_i<E_c\); thus, \(a_i\) is non-constant and \(\operatorname{Cov}_g(\varepsilon,a)>0\). The degenerate case is a group containing only the fixed-temperature segment, or more generally one for which \(a_i\) is constant over all populated states. For the non-degenerate two-segment groups considered here, Eq.~\eqref{eq:positive_temperature_derivative} follows from Eq.~\eqref{eq:energy_derivative_covariance}.

\begin{equation}
    \label{eq:positive_temperature_derivative}
    \frac{\partial e_g}{\partial\beta}<0,
    \qquad
    \frac{\partial e_g}{\partial T_g}
    =
    \frac{\operatorname{Cov}_g(\varepsilon,a)}
    {k_B T_g^2}
    >0.
\end{equation}

The group mean energy is therefore a strictly increasing function of the dynamic closure temperature \(T_g\). Consequently, the energy-consistency relation in Eq.~\eqref{eq:group_reconstruction} admits a unique closure temperature for every admissible group mean energy.

\section{Representative Group-Resolved Channel Profiles}
\label{app:group_resolved_channel_profiles}

 The figures compare the population-weighted group quantities obtained from the database calculations with their reduced representations. The rate panel shows the effective rate coefficient \(\bar{k}_{\rho}\). The remaining panels show the signed group-energy contributions \(\Delta\varepsilon^{\mathrm{vib}}_{r,\rho}\) and the electron-energy loss \(Q^{\mathrm{eff}}_{e,\rho}\). Here, \(\mathrm{H}_2(V_{G_1})\) and \(\mathrm{H}_2(V_{G_2})\) denote molecules whose vibrational states belong to groups \(G_1\) and \(G_2\), respectively. Each rate profile is represented as a function of its selected coordinate, \(\langle\varepsilon_e\rangle\), \(T_v^{G_1}\), or \(T_v^{G_2}\), as indicated on the abscissa. Gray curves show the quantities evaluated over representative VDF conditions. The bold magenta curve shows the reduced one-dimensional rate representation at the nominal VDF condition. Horizontal dashed lines denote the scalar energy terms used in the reduced model. Vertical dashed lines, where present, mark the low-energy cutoff applied to electron-energy-based rate profiles.

\begin{figure}[htbp]
    \centering
    \includegraphics[width=\textwidth]{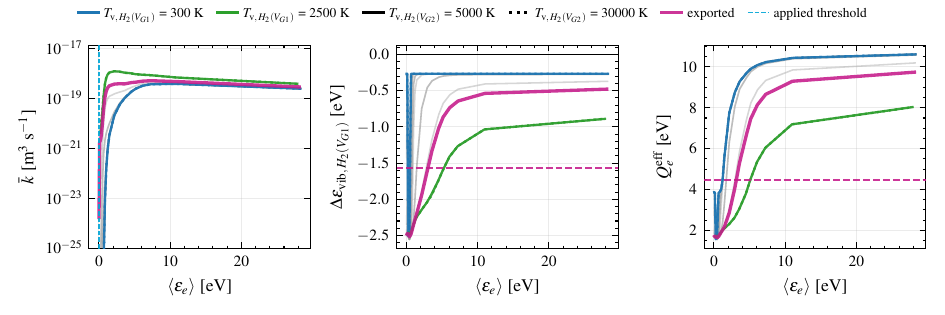}
    \caption{Resonant dissociative electron attachment for \(\mathrm{H}_2(V_{G_1})\), \(e+\mathrm{H}_2(V_{G_1})\rightarrow\mathrm{H}^-+\mathrm{H}\).}
    \label{fig:dea_g1}
\end{figure}

\begin{figure}[htbp]
    \centering
    \includegraphics[width=\textwidth]{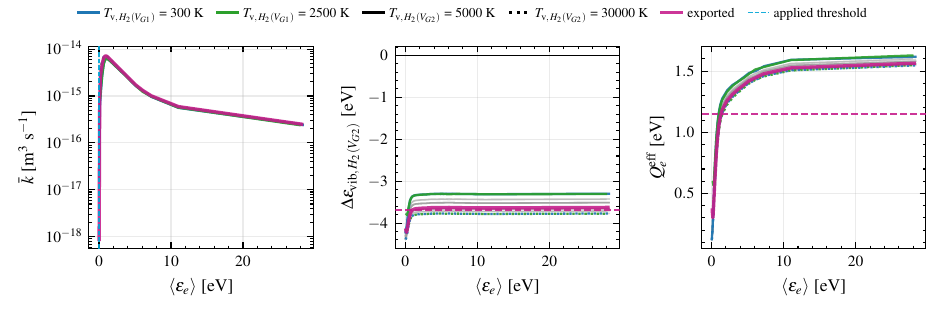}
    \caption{Resonant dissociative electron attachment for \(\mathrm{H}_2(V_{G_2})\), \(e+\mathrm{H}_2(V_{G_2})\rightarrow\mathrm{H}^-+\mathrm{H}\).}
    \label{fig:dea_g2}
\end{figure}

\begin{figure}[htbp]
    \centering
    \includegraphics[width=\textwidth]{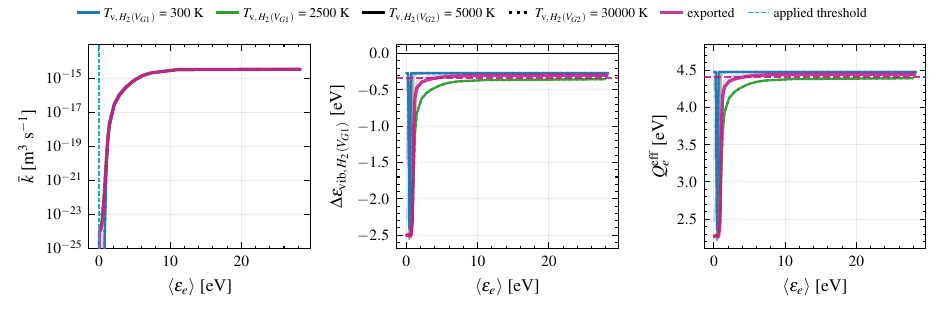}
    \caption{Direct electron-impact dissociation through the \(b^3\Sigma_u^+\) repulsive state for \(\mathrm{H}_2(V_{G_1})\), \(e+\mathrm{H}_2(V_{G_1})\rightarrow e+\mathrm{H}+\mathrm{H}\).}
    \label{fig:exdiss_g1}
\end{figure}

\begin{figure}[htbp]
    \centering
    \includegraphics[width=\textwidth]{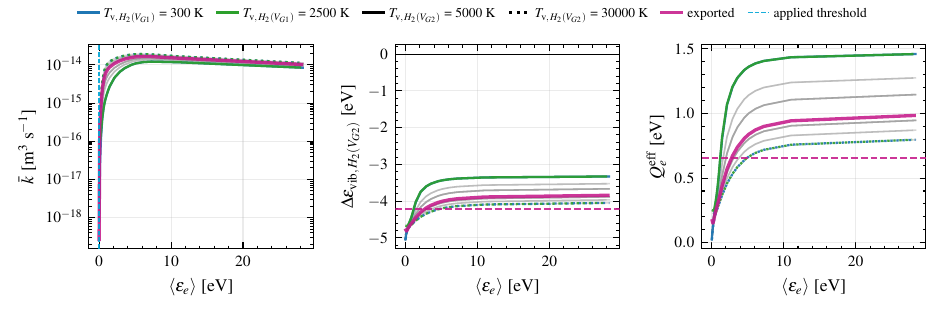}
    \caption{Direct electron-impact dissociation through the \(b^3\Sigma_u^+\) repulsive state for \(\mathrm{H}_2(V_{G_2})\), \(e+\mathrm{H}_2(V_{G_2})\rightarrow e+\mathrm{H}+\mathrm{H}\).}
    \label{fig:exdiss_g2}
\end{figure}

\begin{figure}[htbp]
    \centering
    \includegraphics[width=\textwidth]{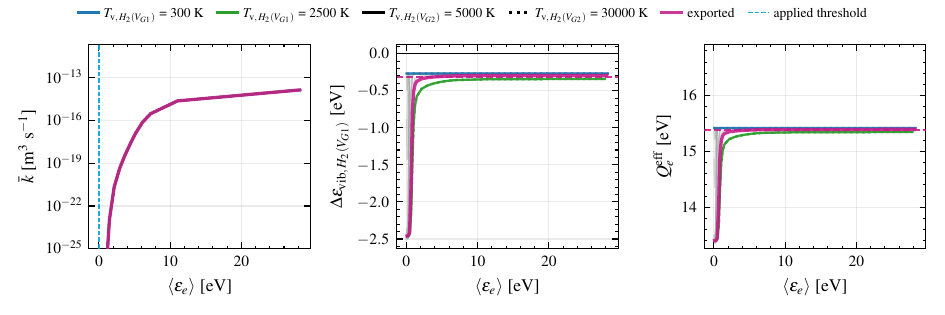}
    \caption{Direct ionization for \(\mathrm{H}_2(V_{G_1})\), \(e+\mathrm{H}_2(V_{G_1})\rightarrow\mathrm{H}_2^++2e\).}
    \label{fig:ioniz_g1}
\end{figure}

\begin{figure}[htbp]
    \centering
    \includegraphics[width=\textwidth]{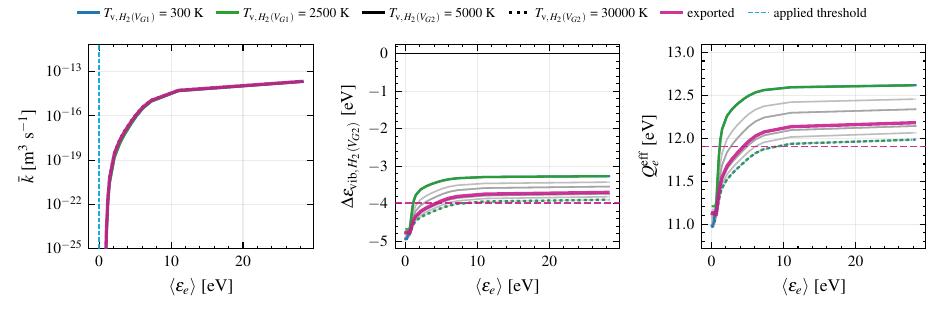}
    \caption{Direct ionization for \(\mathrm{H}_2(V_{G_2})\), \(e+\mathrm{H}_2(V_{G_2})\rightarrow\mathrm{H}_2^++2e\).}
    \label{fig:ioniz_g2}
\end{figure}

\begin{figure}[htbp]
    \centering
    \includegraphics[width=\textwidth]{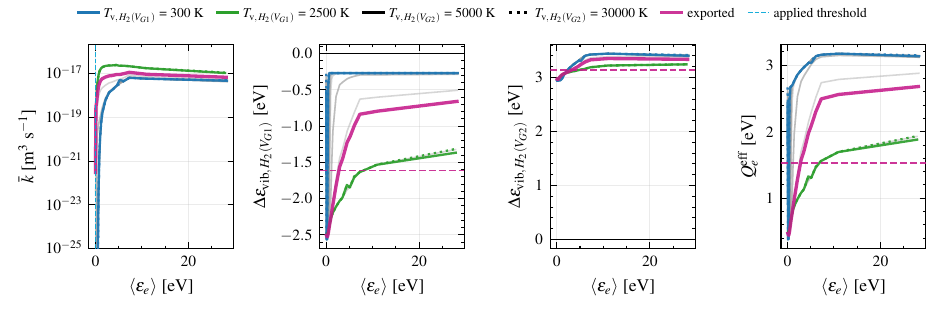}
    \caption{Resonant electron-vibrational excitation, \(e+\mathrm{H}_2(V_{G_1})\rightarrow e+\mathrm{H}_2(V_{G_2})\).}
    \label{fig:exvib_g1g2_exc}
\end{figure}

\begin{figure}[htbp]
    \centering
    \includegraphics[width=\textwidth]{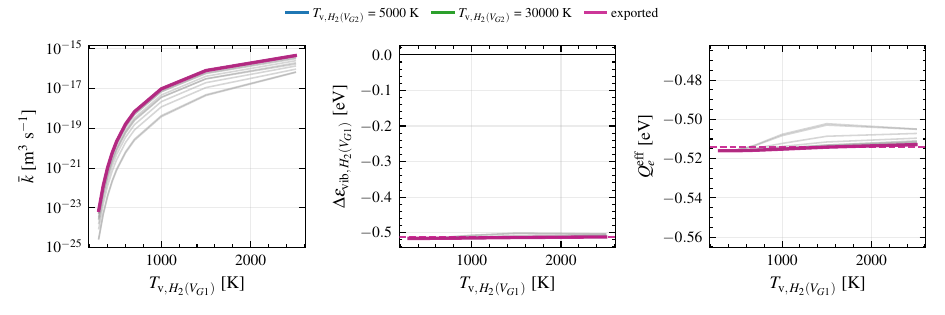}
    \caption{Resonant electron-vibrational de-excitation within \(\mathrm{H}_2(V_{G_1})\), represented as a function of \(T_v^{G_1}\).}
    \label{fig:exvib_g1g1_dexc}
\end{figure}

\clearpage
\bibliographystyle{unsrt}
\bibliography{UT_research}

\end{document}